\documentclass[10pt,journal,compsoc]{IEEEtran}
\ifCLASSOPTIONcompsoc
  \usepackage[nocompress]{cite}
\else
  \usepackage{cite}
\fi
\ifCLASSINFOpdf
   \usepackage[pdftex]{graphicx}
   \DeclareGraphicsExtensions{.pdf,.jpeg,.png}
\else
\fi
\usepackage{amsmath}
\begin{document}
%
\title{Intrusion Alert Prediction Using a Hidden Markov Model}
%
%
%
%

\author{Udaya Sampath K. Perera~Miriya Thanthrige, 
        Jagath~Samarabandu 
        and~Xianbin~Wang 

\thanks{This work was supported by the Natural Sciences and Engineering Research Council of Canada (NSERC).
The authors are with Department of Electrical and Computer Engineering, University of Western Ontario, London, Ontario, Canada.
Email:\{mperer4, jagath, xianbin.wang\}@uwo.ca}
}
\author{Udaya Sampath K. Perera~Miriya Thanthrige, 
        Jagath~Samarabandu 
        and~Xianbin~Wang 
\IEEEcompsocitemizethanks{\IEEEcompsocthanksitem This work was supported by the Natural Sciences and Engineering Research Council of Canada (NSERC).
The authors are with Department of Electrical and Computer Engineering, University of Western Ontario, London, Ontario, Canada.\protect\\
E-mail: \{mperer4, jagath, xianbin.wang\}@uwo.ca}
}

%
%

\markboth{Intrusion Alert Prediction Using a Hidden Markov Model}%
{Shell \MakeLowercase{\textit{et al.}}: Intrusion Alert Prediction Using a Hidden Markov Model}

%



\IEEEtitleabstractindextext{%
\begin{abstract}
Intrusion detection is only a starting step in securing IT infrastructure. Prediction of intrusions is the next step to provide an active defense against incoming attacks. Current intrusion prediction methods focus mainly on prediction of either intrusion type or intrusion category and do not use or provide contextual information such as source and target IP address. In addition most of them are dependant on domain knowledge and specific scenario knowledge. The proposed algorithm employs a bag-of-words model together with a hidden Markov model which not depend on specific domain knowledge. Since this algorithm depends on a training process it  is adaptable to different conditions. A key advantage of the proposed algorithm is the inclusion of contextual data such as source IP address, destination IP
range, alert type and alert category in its prediction, which is crucial for an eventual response.  Experiments conducted using a public data set generated over 2500 alert predictions and achieved accuracy of 81\% and 77\% for single step and five step predictions respectively for prediction of the next alert cluster. It also achieved an accuracy of prediction of 95\% and 92\% for single step and five step predictions respectively for prediction of the next alert category. The proposed methods achieved a prediction accuracy improvement of 5\% for alert category over existing  variable length Markov chain intrusion prediction methods, while providing more information for a possible defense.
\end{abstract}

}

\maketitle

\IEEEdisplaynontitleabstractindextext

%
\IEEEpeerreviewmaketitle

\IEEEraisesectionheading{\section{Introduction}\label{sec:introduction}}

%
%
%
%
\IEEEPARstart{H}{uman} activities associated with computational devices are rapidly growing day by day. Nowadays most smart devices are capable of connecting to the internet. The world is becoming a more connected place and people are heavily relying on services offered based on computer networks. With such a high involvement with interconnected devices, sensitive and valuable information are rapidly exchanging between computer networks and devices. Initially, most of the research activities were focused on intrusion detection methods. Such intrusion detection systems identify the attacks that have already happened or currently happening in the network. To improve network security, advanced system models are required, which are able to detect impending or ongoing intrusions and take countermeasures to stop intrusion activities. Researchers are focusing on developing sophisticated Intrusion Response System (IRS) to fulfill this requirement.\\
Prediction of future intrusion activities plays a key role in intrusion response, which helps in identifying and executing response actions before an intrusion occurs. Intrusion detection systems (IDSs) generate intrusion alerts once they detect network intrusions. These intrusion alerts can then be used to understand attacker strategies and predict future network intrusions.\\
Most of the previous intrusion prediction methods were mainly focused on prediction of either alert type or alert category (i.e. identification of next stage attack) \cite{5963162}. A proper response action cannot be executed by only knowing future intrusion type or intrusion category. In order to execute useful response action, it is important to predict other important parameters such as the attacker and the victim of the attack. A response action can then be executed on a specific network segment without  unduly affecting the whole network. In addition response actions may be targeted specifically to the attacker.\\
In this paper, we propose A Hidden Markov model based algorithm for predicting the next intrusion alert, given a sequence of intrusion alerts from an existing IDS tool such as "Snort". Our Proposed alert prediction method is based on prediction of the next alert cluster, which contains source IP address, destination IP range, alert type and alert category. Hence, prediction of next alert cluster provides more information about future strategies of the attacker. Also, Our Proposed alert prediction method does not depend on specific domain knowledge. Instead, it depends on a training process. Hence the proposed algorithm is adaptable to different conditions.\\
The rest of the paper is organized as follows. Section II describes related work on intrusion prediction. Section III describes architecture of the proposed alert prediction framework, which includes the Bag of Words (BoW) model based k-means alert clustering process and implementation of HMM for alert prediction. Section IV presents performance evaluation of proposed alert prediction method with respect to the number of clusters and HMM parameters. It also compares prediction accuracy of alert cluster based sequence modeling and alert category based sequence modeling. Section V includes critical analysis of results and challenges involve in the alert prediction process.

 
\hfill October 23, 2016

\section{Related Work}

Intrusion prediction can be mainly categorized as network attack graph based prediction \cite{li2007data}, \cite{cheng2011novel}, \cite{wang2006using} and predictions based on sequence modeling techniques (such as Markov Model, Hidden Markov Model, Bayesian Networks and Dynamic Programming) \cite{fava2008projecting}, \cite{du2010toward}, \cite{7387905}. Hidden Markov Model (HMM) is one of the widely used sequential data modeling methods which was introduced in the late 1960s by Baum and his colleagues \cite{baum1966statistical}, \cite{baum1967inequality}. Due to the rich mathematical structure of HMM, it widely applied in real world applications such as speech recognition, handwriting pattern recognition, gesture recognition, intrusion detection and speech tagging.\\
Intrusion alerts typically contain two fields that are used to identify an intrusion which caused the corresponding intrusion alert. These fields are called ``alert type" and ``alert category." Alert type contains detailed information about an intrusion whereas alert category contains higher level information about an intrusion. Most of the previous intrusion prediction methods were mainly focused on prediction of either alert type or alert category. A proper response action cannot be executed by only knowing future intrusion type or intrusion category.\\
Qin and Lee proposed attack projection scheme based on probabilistic reasoning method \cite{qin2004attack}. They used attack tree to develop Bayesian network. The main limitation of this process is prediction depends on the availability of an attack plan library. Also, it requires mapping best possible attack graph for a given observation scenario.\\
Similar to the method proposed by Qin and Lee, Ramki et al. proposed Bayesian network based directed acyclic graph (DAG) aided alert prediction method \cite{ramaki2015real} where they evaluated the correlation between alert types. Based on correlation and generated attack graph, they predicted next possible alert type that can occur. They used the DARPA2000 dataset \cite{DARPA} for their alert type prediction.\\
Li et al. \cite{li2007data} proposed attack graph based intrusion prediction method. They used association rule mining to generate attack graphs. By observing the attack graph, it is possible to predict possible future steps once several initial steps are observed. They used the DARPA1999 \cite{DARPA} and DARPA2000 \cite{DARPA} datasets for their experiments.\\
Multi-stage attack forecasts based on probabilistic matching was proposed by Cheng et al. where prediction of attack steps were done by measuring the difference between the stored and the actual multi-stage attack session graphs (ASG) \cite{5963162}. Their method was inspired by the generalized Hough transform and polygonal curves mapping concept. Probabilistic mapping does not require exact mapping, therefore, this method performs better when it compare with exact mapping techniques such as Longest Common Sub-sequence (LCS) algorithm. The major limitation of this method is that prediction depends on the availability of an attack graph library.\\ 
Lippmann et al. analyzed attack graph generation methods proposed by various researchers and illustrated that most of these graph generation involved network with less than 20 nodes and most of them had poor capability of scaling \cite{lippmann2005annotated}. This finding indicates that intrusion activity prediction methods based on attack graphs are not very feasible to employ in practice due to scaling issues.\\
Fava et al. \cite{fava2008projecting} proposed Cyberattacks projecting method using variable-length markov models. They have selected three alert attributes namely alert category, alert description and destination IP address of an alert as symbols in the Markov model and generated three different Markov models for alert category, alert description and destination IP address. Using these models, they predicted the next alert category, next alert description and destination IP address separately.\\
One main limitation of their method is that they are predicting alert type, alert category and destination IP separately. So there are possibilities that combination of these three may not be a valid combination. As an example, the predicted alert category is ``Intrusion Root" and predicted alert description is ``ICMP PING NMAP". However, predicted alert description does not belong to the predicted alert category. To address this limitation, Du et al. \cite{du2010toward} presented an ensemble techniques based alert attribute combination method. Based on ensemble techniques, they have a proposed method to convert VMM alert prediction output to a score with respect to the host of the network. This score is used to predict next possible host that can be attacked. Major limitation of their method is that they have only interested to predict next host without predicting attack type.\\
Our proposed alert prediction framework is different from existing approaches in following aspects: It does not depend on specific domain knowledge. Instead, it depends on a training process. Hence the proposed algorithm is adaptable to different conditions. Also, it is based on prediction of the next alert cluster, which contains source IP address, destination IP range, alert type and alert category. Hence, prediction of next alert cluster provides more information about future strategies of the attacker.

\section{System Architecture}

\begin{figure}[!t]
\centering
\includegraphics[width=2.5in]{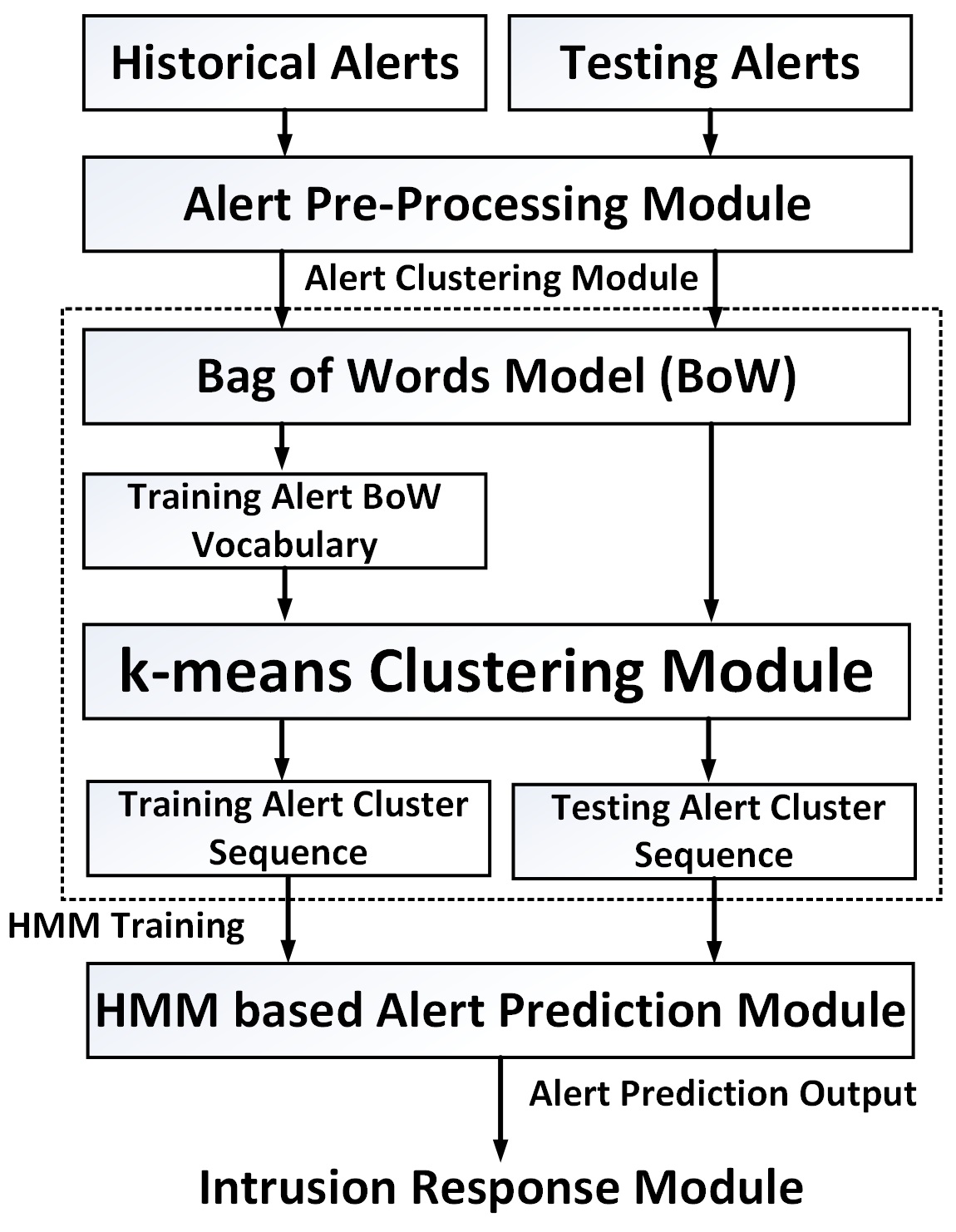}
\caption{Proposed hidden Markov model based alert prediction module.}
\label{proposedsysarchi}
\end{figure}

The proposed alert prediction module consists of three main modules called alert pre-processing module, alert clustering module and alert prediction module as shown in Figure \ref{proposedsysarchi}. Output of the alert prediction module is forwarded onto intrusion response module.
In this paper, sequential data modeling concept was employed to predict next possible intrusion alert for a given historical alert sequence. HMM was used as sequential data modeling method. To build a sequential data series in network security domain, alert clustering concept was used. Intrusion alerts were clustered based on their feature similarities and then for a given sequence of alerts, alert clustering module produced a sequence of clusters. Those sequence of clusters were considered to build a sequential data series in our model. The HMM is used to predict next cluster using the sequence data series which was built in previous step. Since predicted cluster represents certain characteristics of alert attributes, that information can be used by intrusion response module to select and execute relevant response actions to mitigate future intrusions.

\subsection{Alert Pre-Processing and Clustering} \label{alertcluster}

Intrusion alerts generated by intrusion detection systems (IDSs) was utilized by the proposed alert prediction framework. Snort was used as an intrusion detection system \cite{4snort} for this work. Snort is an open-source network intrusion detection system (NIDS) \cite{4snort} . A Snort alert consists two major fields that are used to identify an intrusion which caused the corresponding intrusion alert. These fields are called ``alert type" (also known as alert description or alert signature) and ``alert category" (also known as alert signature class). The first step of the pre-processing was to format raw alerts from different IDSs to a common format and then duplicated alerts were removed using selected features. Timestamp, source IP, source port, destination IP, destination port, alert signature and alert signature class were used to identify duplicate alerts. The main objective of alert clustering process is to generate a sequence of symbols that can be modeled using sequential data modeling. Bag of Words (BoW) model, which is a popular concept in text document classification and k-means clustering algorithm were used to cluster alerts.

\subsubsection{Bag of Words (BoW) Model }  \label{bofwordvocabulary}
Bag of Words (BoW) model is a popular document analysis algorithm used in text and image classification. Few sample alerts created by Snort is shown in table \ref{snortalertdarpa} were used to illustrate this process. In the first step, important attributes of alert were selected to develop vocabulary. Source IP address, source IP port, destination IP address, destination IP port, signature (alert type) and signature class (alert category) were selected to generate vocabulary. Each unique word in alert attributes was added to the vocabulary. IP address was treated in a special way where four segments of an address was considered as four separate words. BoW vocabulary which was generated from sample alerts shown in table \ref{snortalertdarpa} is shown in table \ref{BowVocabularyforsamplealerts}.
\begin{table}[!t]
\centering
\caption{Few Snort alerts.}
\resizebox{\columnwidth}{!}{%
\begin{tabular}{ | l | l | l | l | }
\hline
	\textbf{Source\,IP} & \textbf{Destination\,IP} & \textbf{Alert\,Type\,ID} & \textbf{Alert\,Category} \\ \hline
	172.16.112.100 & 172.16.112.20 & 650 & attempted-recon \\ \hline
	172.16.112.50 & 172.16.113.169 & 506 & sdf \\ \hline
	172.16.116.201 & 206.83.105.134 & 36489 & trojan-activity \\ \hline
	207.25.71.186 & 172.16.117.132 & 684 & unknown \\ \hline
	209.87.178.183 & 192.168.5.122 & 41297 & web-application-attack \\ \hline
\end{tabular}
}
\label{snortalertdarpa}
\end{table}
\begin{table}[!t]
\centering
\caption{BoW vocabulary generated from alerts shown in table \ref{snortalertdarpa}. }
\resizebox{\columnwidth}{!}{%
\begin{tabular}{ | l | l | l | l | l | l | l | l | l | }
\hline
	\multicolumn{9}{|c|}{\textbf{Bag of Words Vocabulary}}\\ \hline
	activity &  application &  attack &  attempted &  recon &  sdf &  trojan & unknown & web \\ \hline
	16 & 168 & 169 & 172 & 178 & 183 & 186 & 192 & 50 \\ \hline
	83 & 87 & 105 & 112 & 113 & 115 & 116 & 117 & \  \\ \hline
	684 & 71 & 207 & 209 & 25 & 506 & 650 & 20 & \  \\ \hline
	206 & 134 &  36489 & 201  & 100 & 122 & 132 & 41297 & \  \\ \hline
\end{tabular}
}
\label{BowVocabularyforsamplealerts}
\end{table}
\noindent The next step is to cluster alerts based on their attributes. For the clustering process, k-means clustering algorithm was used. The k-means algorithm was first used by James MacQueen in 1967 \cite{macqueen1967some}. Historical alerts were used to generate set of alerts clusters. Once new alert was received, it was assigned to the best matching cluster using k-means algorithm.\\
The main objective of clustering is to combine alert attributes and generates a single unit that represents different alerts attributes. This single unit is the cluster ID. A Cluster ID sequence for given input alert sequence was generated by the alert clustering module and that sequence was used as a sequential data series in our model. This sequential data series was used to build HMM module and predict future alert clusters.\\
Let k be the number of clusters was generated by k-means algorithm (represent cluster set C = \{$C_1, C_2,...,C_i,...,C_k$\}). Furthermore, T number of alerts ordered based on their time of origin (timestamp) was represented as $A_1, A_2,...,A_i,...,A_k$. The best matching cluster (from cluster set C) for each alert was assigned by the alert clustering module. To illustrate alert clustering process, total number of clusters was selected as five and then vocabulary (which was generated in previous step) was clustered to five clusters by the k-means clustering module. These clusters are shown in table \ref{BowVocabularyclusterforsamplealerts}.
\begin{table}[!t]
\centering
\caption{BoW vocabulary clusters generated from alerts shown in table \ref{snortalertdarpa}. }
\resizebox{\columnwidth}{!}{%
\begin{tabular}{ | l | l|  l|  l  |l | l  |l|  l | l | l | l | }
\hline
  \textbf{Cluster Name} & \multicolumn{10}{c|}{\textbf{Elements in Clusters}} \\ \hline
	Cluster 0 & 87 & 183 & 178 & 209 & 41297 & 168 & 192 & 122 &  web&  application  \\ \hline
	Cluster 1 & 172 & 16  & 112 & 113 & 169 & 50 & 506 & 178 & 168&  sdf \\ \hline
	Cluster 2 & 206  & 172 & 36489 & 16 & 134 & 83 & 201 & 116 & 105&  trojan \\ \hline
	Cluster 3 & 71 & 132  & 207 & 25 & 172 & 16 & 684 & 186 & 117&  unknown \\ \hline
	Cluster 4 & 16 & 172 & 100 & 112  & 115 & 20 & 650 & 186 &  recon &  attempted \\ \hline
\end{tabular}
}
\label{BowVocabularyclusterforsamplealerts}
\end{table}
Based on these observations, it can be seen that cluster 0 represents ``web application attack" between IP address 209.87.178.183 and IP address 192.168.5.122, cluster 2 represents ``trojan activity attack" between IP address 172.16.116.201 and IP address 206.83.105.134 and so on. When a new alert was received, the best matching cluster for that alert was assigned by the alert clustering module based on alert attributes.
\subsection{Building Hidden Markov Models} \label{buildHMM}
A hidden Markov model was used to model sequential data produced from alert clustering module. HMM consist of hidden states which produced the observations. Alert cluster sequence was considered as an observation sequence in this model.
Parameters of hidden Markov model is defined as:
\begin{enumerate}
\item{${\bf N}$, Number of hidden states in the HMM.}
Where individual states are denoted as ${\bf S} = \{S_1, S_2, \ldots , S_N\}$ and state at time t is denoted as $q_t$.
\item{${\bf M}$, Number of distinct observation symbols per state.}
Where individual symbols are denoted as ${\bf V} = \{v_1, v_2, \ldots , v_M\}. $
\item{State transition probability ({\bf A}) $N$x$N$ matrix.}
Where $a_{ij}$ represents the state transition probability form state $i$ to state $j$.\\
$a_{ij} =P(q_{t+1}=S_j|q_t=S_i), where \, 1\le i \le N and 1\le j \le N$.
\item{Observation emission probability ({\bf B})  $N$x$M$ matrix.}
Where $k^{th}$ observation emission probability of the state $j$ is represented by ${b_j(k)}$.\\
${b_j(k)}=P(v_{k}$ $ at$ $t|q_t=S_j), where \, 1\le j \le N, 1\le k \le M$.
\item{Initial state probability distribution (${\pi}$).}
Where ${\pi}$ represents the initial states probabilities of the HMM.\\
${\pi_i}=P(q_1=S_i) , 1 \le i \le N$.
\item{Observation sequence ({\bf O}).}
The observation sequence of length T is represented as ${\bf O} = O_1, O_2, \ldots , O_t,\ldots ,O_T$, Where $O_t$ is one of observation symbol from ${\bf V}$.\\
\end{enumerate}
In our model, observations are distinct alert cluster IDs generated from alert clustering module. The cluster ID sequence generated by historical alerts was used to train the HMM. HMM was trained using Baum-Welch algorithm \cite{18626}.

\subsubsection{Hidden Markov Model Based Sequence Prediction}
Trained HMM was used to predict next alert clusters based on observed alert cluster sequence.\\
{\bf Notation:}
\begin{itemize}
 	\item {Trained HMM ${\lambda}=({\bf A},\, {\bf B},\, {\pi})$.}
	\item {${\bf T}$ = Length of observed testing alert sequence.}
	\item {${\bf k}$ = Number of clusters generated by k-means clustering module.}
	\item {Cluster set generated by k-means $ {\bf C} = \{C_1, C_2,...,C_t,...,C_k$\}.}
	\item {Observed testing alert sequence ${\delta}$ = $ A_1, A_2, \ldots , A_t,\ldots ,A_T. $}
	\item {Corresponding cluster sequence for observed testing alert sequence ${\omega}  = {C^A}_1, {C^A}_2, \ldots , {C^A}_t,\ldots ,{C^A}_T$ where ${C^A}_t \in  C  \quad for\quad 1 \, \le \, t \, \le \, T. $}
	\item {Observation sequence {\bf O} = $O_1, O_2, \ldots , O_t,\ldots ,O_T.$}
\end{itemize}
In this model, observed alert cluster sequence was taken as observation sequence.\\
Hence, {\bf O} = ${\omega} = {C^A}_1, {C^A}_2, \ldots , {C^A}_t,\ldots ,{C^A}_T.$\\
The best hidden state sequence for the observed testing cluster ID sequence (${\omega} = {C^A}_1, {C^A}_2, \ldots , {C^A}_t,\ldots ,{C^A}_T$) was identified using Viterbi algorithm.\\
Let ${\bf Q} = q_1, q_2, \ldots , q_t,\ldots ,q_T$ is the best hidden alert sequence. Using state transition probability matrix ({\bf A}) and observation emission probability matrix ({\bf B}), probability of each cluster for the next position is calculated. Based on the probabilities of clusters, the best possible candidate for next cluster is selected. Let last element of ${\bf Q}$ was $S_j$. Probability of cluster $C_i$ to be in next observation ($P_{T+1}\,(C_i))$ can be calculated using equation \ref{eqn:HMMprobpre}:
\begin{equation} \label{eqn:HMMprobpre}
\begin{split}
\textrm{ $P_{T+1}\,(C_i)$ } =\, \sum_{r=1}^{N} a_{jr}\,b_r(i) \text{.} \\
\textrm{ Where \, $C_i$} \in  C \text{.}
\end{split}
\end{equation}
Using equation \ref{eqn:HMMprobpre}, probability of each cluster to be in next position was calculated. In this model, three possible clusters were identified for next observation based on their probability values. These three were labeled as level 1, 2 and 3 prediction. Level 1 prediction includes cluster which has the highest probability. Level 2 prediction includes highest and second highest probability clusters. Level 3 prediction includes first, second and third highest probability clusters.\\
For multiple step prediction (prediction of many data points in one prediction step), alert cluster which was predicted in previous prediction stage was appended to input observation sequence when predicting next point. 

\section{Experimental Results}
The DARPA (Defense Advanced Research Projects Agency) \cite{DARPA} dataset was used in the experimental process. It consists of two multi-stage attacks labeled LLDOS 1.0 and LLDOS 2.0.2. Both these attacks have five main stages \cite{DARPA}. Alert pre-processing and Hidden Markov model based alert prediction modules were implemented using python programming language. Snort IDS (version 2.9.8.0 with rule set 2980) was used for intrusion alert generation. MYSQL database was used as a database for the platform. Bag of words model implementation and k-means clustering implementation was done using python scikit-learn library \cite{scikit-learn}.\\
DARPA 2000 raw network packets were sent into Snort intrusion detection system to generate alerts. 11,264 and 10,468 raw alerts were generated by LLDOS 1.0 and LLDOS 2.0.2 respectively. These alerts were then sent into the alert pre-processing module. Redundant alerts with same attributes were filtered by the alert pre-processing module. After this process the total number of alerts were reduced to 5113 and 5645 respectively. During the next step, these alerts were forwarded to the alert clustering module.\\ 
9 alerts categories and 21 alerts descriptions were produced for LLDOS 1.0 while 9 alerts categories and 19 alerts descriptions were produced for LLDOS 2.0.2 by Snort IDS. DARPA 2000 LLDOS 1.0 was used for vocabulary generation and then that vocabulary was clustered using the k-means algorithm. There were 243 unique words in the vocabulary. The DARPA 2000 LLDOS 2.0.2 was used as a test network that must be monitored and predicted future activities.

\subsection{Alert Prediction} \label{cha:AlertPre}

Performance evaluation of HMM based alert prediction module was performed using the alert cluster ID sequence generated by DARPA 2000 LLDOS 2.0.2 dataset. The length of the cluster ID sequence generated for DARPA 2000 LLDOS 2.0.2 was 5645. 2500 data points were used to train HMM and other segment was used to evaluate prediction capabilities of the model.\\
In order to easily compare our results with existing alert prediction research activities, three accuracy values were calculated as shown in below.
\begin{enumerate}
	\item {Level 1 prediction accuracy (${\alpha L}^1$)}.	In level 1 prediction accuracy, a correct prediction means that, most probable prediction of the alert prediction framework is matched with next action of the attacker. 
	\item {Level 2 prediction accuracy (${\alpha L}^2$)}.	In level 2 prediction accuracy, a correct prediction means that, one of two most probable predictions of the alert prediction framework is matched with next action of the attacker. 
	\item {Level 3 prediction accuracy (${\alpha L}^3$)}. In level 3 prediction accuracy, a correct prediction means that, one of three most probable predictions of alert prediction framework is matched with next action of the attacker. 
\end{enumerate}
Following experiments were conducted to evaluate performance of proposed alert prediction framework.\\
\subsubsection{Performance Evaluation of Next Alert Cluster Prediction\\} \label{penac}

\begin{itemize}
	\item {Effect of number of hidden states in HMM on the prediction accuracy}
\end{itemize}
                                             
\begin{figure}[!t]
\centering
\includegraphics[width=3.0in]{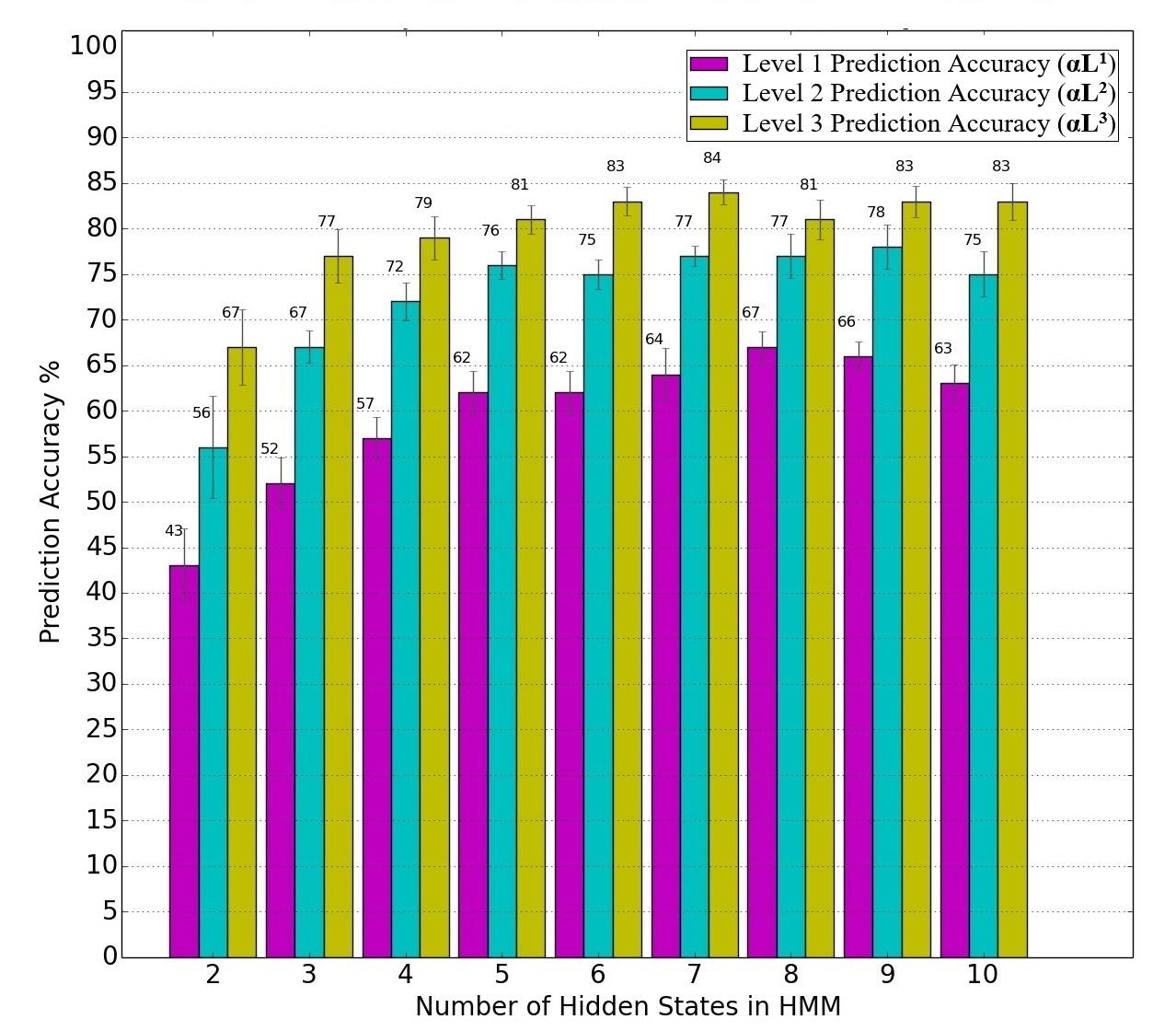}
\caption{Prediction accuracy of alert clusters variation with number of states in HMM.}
\label{result1}
\end{figure}

\noindent There is no straightforward method to determine the best number of hidden states of a HMM \cite{18626}. To find out the best model, number of hidden states were changed from 2 to 10 with increment of 1 at a time. Accuracy of alert prediction are shown in Figure \ref{result1}. The maximum ${\alpha L}^1$ was observed as 67\% for a HMM with 8 hidden states. The lowest ${\alpha L}^1$ was observed as 43\% for a HMM with 2 hidden states. It was observed that, with the increment of the number of states ${\alpha L}^1$ was initially increased and then decreased. The maximum ${\alpha L}^3$ was observed as 84\% for a HMM with 7 hidden states. ${\alpha L}^1$ was changed from 43\% to 67\% with the varying number of hidden states. During the experiments cluster size was kept at 10. A HMM with 8 hidden states was achieved the maximum accuracy of prediction (${\alpha L}^1$). Therefore, that HMM module was selected for other experiments beyond this point.\\

\begin{itemize}
	\item {Effect of the training length on the prediction accuracy}
\end{itemize}

\begin{figure}[!t]
\centering
\includegraphics[width=3.5in]{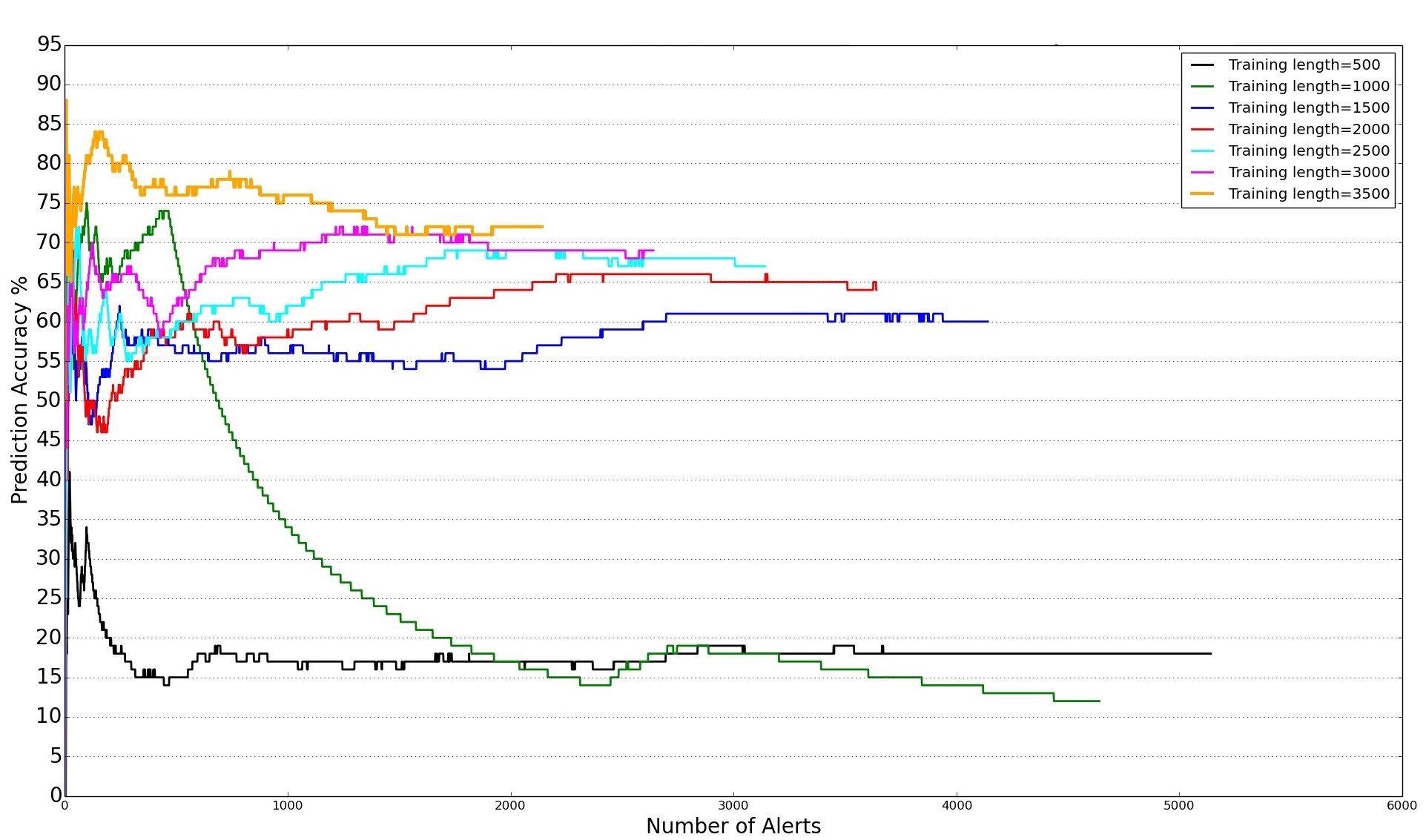}
\caption{Level 1 prediction accuracy (${\alpha L}^1$) of alert clusters variation with length of training sequence.}
\label{result3}
\end{figure}

\noindent Variation of prediction accuracy (${\alpha L}^1$) with the length of the training sequence is shown in Figure \ref{result3}. The length of training sequence was changed from 500 to 3500 with the increments of 500 at a time. The remaining segment was used to evaluate prediction performance. Results indicate that, insufficient training lengths produce very low ${\alpha L}^1$ (around 20\%) for training length 500 and 1000. With the increment of training length ${\alpha L}^1$ was increased and the maximum value for ${\alpha L}^1$ was recorded as 73\% for the training length of 3500.\\

\begin{itemize}
	\item {Effect of number of clusters on the prediction accuracy}
\end{itemize}
	
\begin{figure}[!t]
\centering
\includegraphics[width=3.0in]{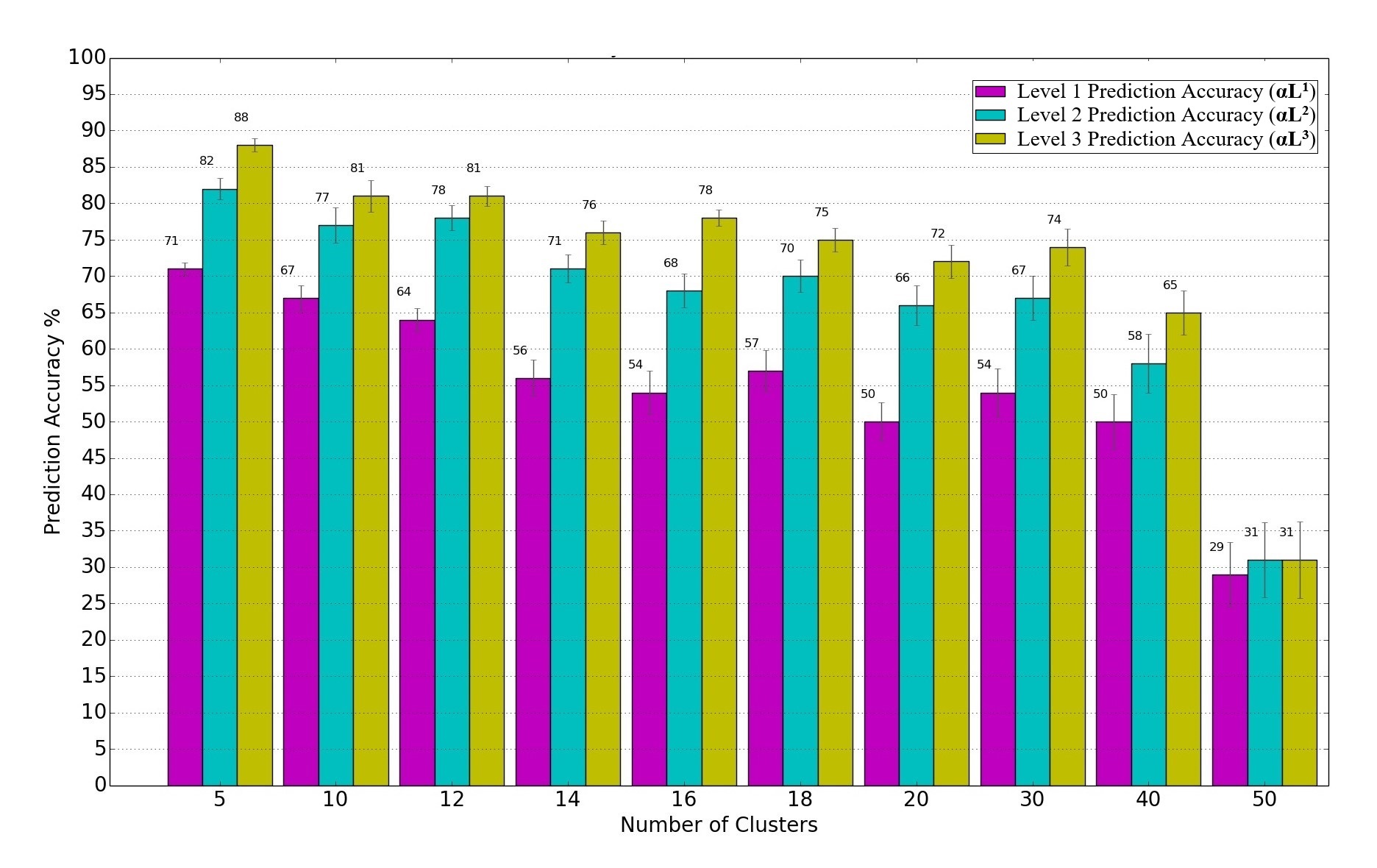}
\caption{Prediction accuracy of alert clusters variation with number of clusters.}
\label{result4}
\end{figure}

\noindent To illustrate the impact on the size of unique observation symbols, number of clusters were changed from 5 to 50. Alert perdition accuracy variation with the number of clusters is shown in Figure \ref{result4}. It is obvious that, a high number of clusters produced better separation among alerts and which result in better grouping. For DARPA dataset there were 9 different alert categories and 19 different alert types produced by Snort. Therefore, in order to group alerts by their corresponding alert category and corresponding alert type at least 9 and 19 clusters were required respectively. The maximum value for ${\alpha L}^1$ was 88\% recorded for 5 clusters and the lowest 31\% was recorded for 50 clusters. However, up to 30 clusters ${\alpha L}^3$ was remained over 70\%. It was observed that HMM prediction fail to achieve better accuracy for higher observation symbol size.\\

\begin{itemize}
	\item {Performance evaluation of multiple intrusions predictions}
\end{itemize}

\begin{figure}[!t]
\centering
\includegraphics[width=3.0in]{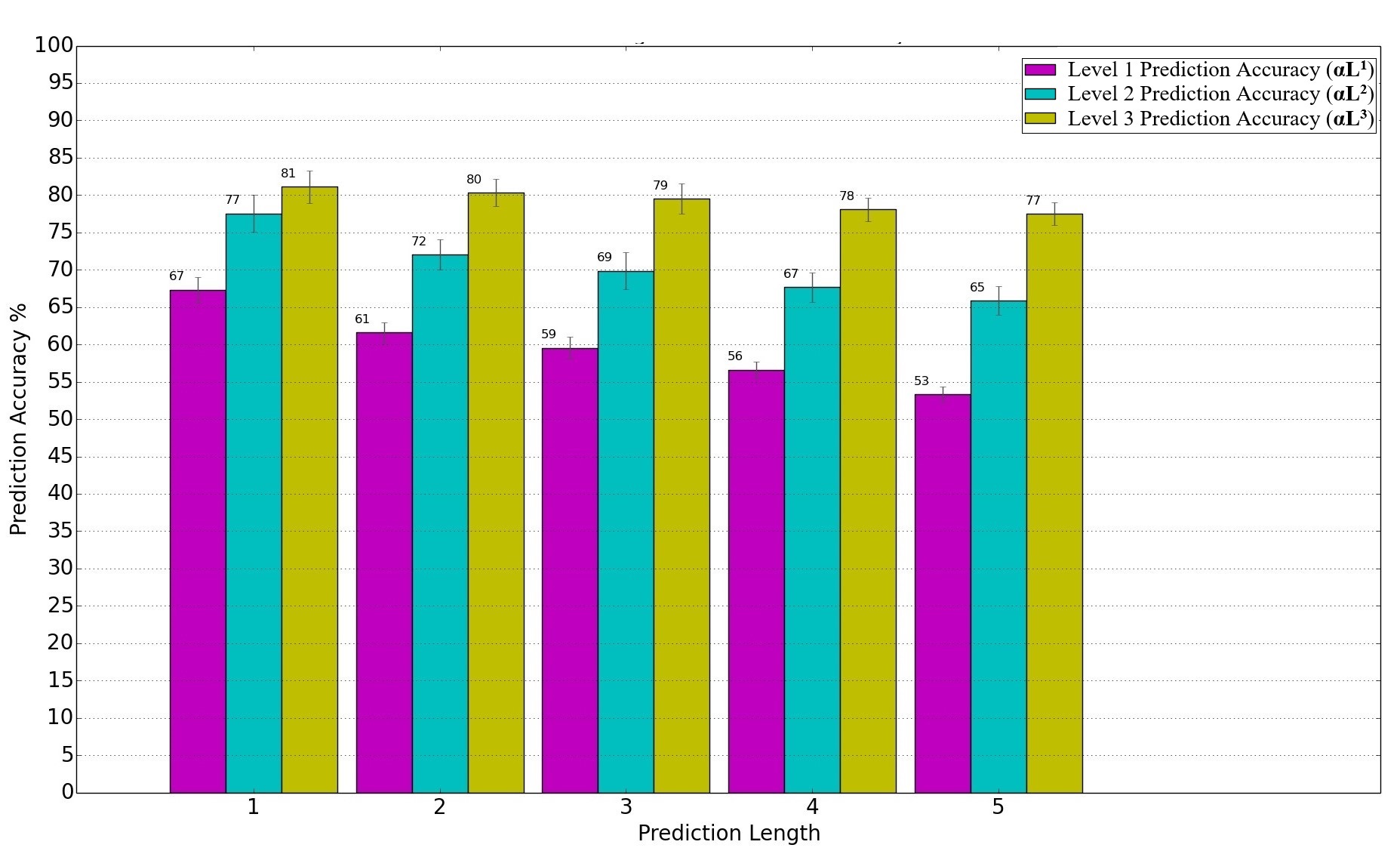}
\caption{Prediction accuracy of alert clusters variation with multiple data point prediction.}
\label{result10}
\end{figure}

\noindent To examine the multiple data points prediction capability of HMM prediction module, the length of prediction ($l$) was changed from 1 to 5 and accuracy of prediction was recorded for 10 clusters. Accuracy of predictions is shown in Figure \ref{result10}. It was observed that, with the increment of prediction length accuracy of prediction was decreased. One data point prediction achieved the highest accuracy of 67\%, 77\% and 81\% for ${\alpha L}^1$, ${\alpha L}^2$ and ${\alpha L}^3$ correspondingly while five data point prediction achieved lowest with 53\%, 65\% and 77\% correspondingly.\\

\subsubsection{Performance Evaluation of Next Alert Category (alert class) Prediction} \label{penacategory}

Most of alert prediction research activities were largely focused on predicting future alert category. In order to compare proposed alert prediction framework with other network intrusion prediction research activities, alert category based prediction was evaluated. For following experiments, sequential data series was generated by considering an alert category as an observation. For DARPA 2000 LLDOS 2.0.2 dataset 9 different alert categories were generated by Snort. These alerts categories were mapped to the symbols. By using this mapping, alert sequence generated for LLDOS 2.0.2 was converted to a symbol series. Similar to the other experiments 2500 data points were used to train HMM and the remaining segment was used to evaluate prediction performance.\\

\begin{figure}[!t]
\centering
\includegraphics[width=3.0in]{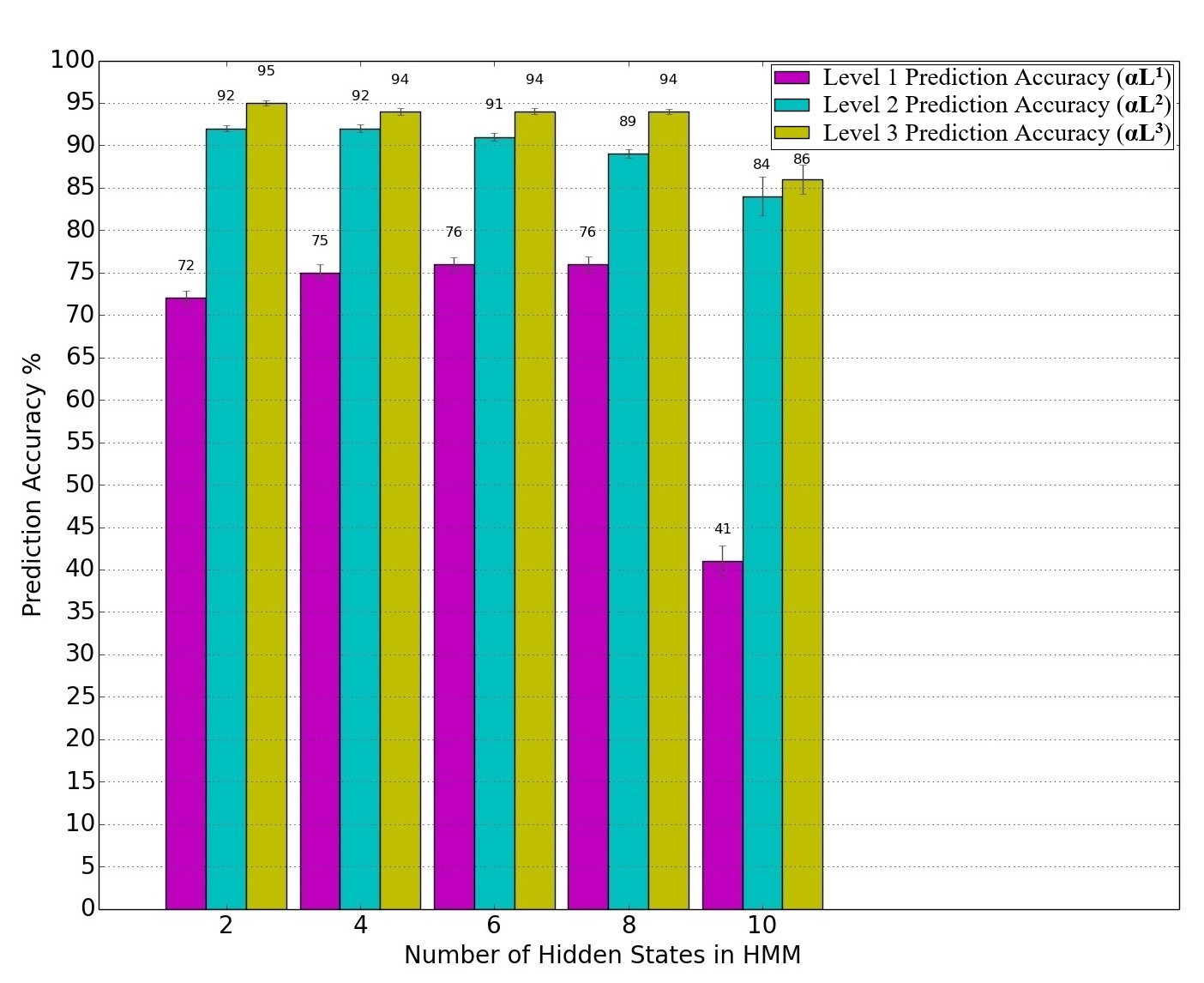}
\caption{Prediction accuracy of alert category variation with number of hidden states in HMM.}
\label{result11}
\end{figure}

\begin{itemize}
	\item {Effect of number of hidden states in HMM on the prediction accuracy}
\end{itemize}
\noindent To evaluate the accuracy of alert prediction with the number of hidden states, the number of hidden states were changed from 2 to 10 with increment of 2 at a time. The results of this experiment is shown in Figure \ref{result11}. The maximum ${\alpha L}^1$ of 76\% was recorded for HMMs with 6 and 8 number of hidden states. Also, it was observed that sudden drop of accuracy was recorded for a HMM with 10 hidden states. It was observed that, for this model number of hidden states were higher than the number of unique observations. Also, 10\% of ${\alpha L}^3$ accuracy improvement was observed for prediction of alert categories compared to prediction of alert clusters size of 10.\\

\begin{itemize}
	\item {Performance evaluation of multiple intrusions predictions}
\end{itemize}

\begin{figure}[!t]
\centering
\includegraphics[width=3.0in]{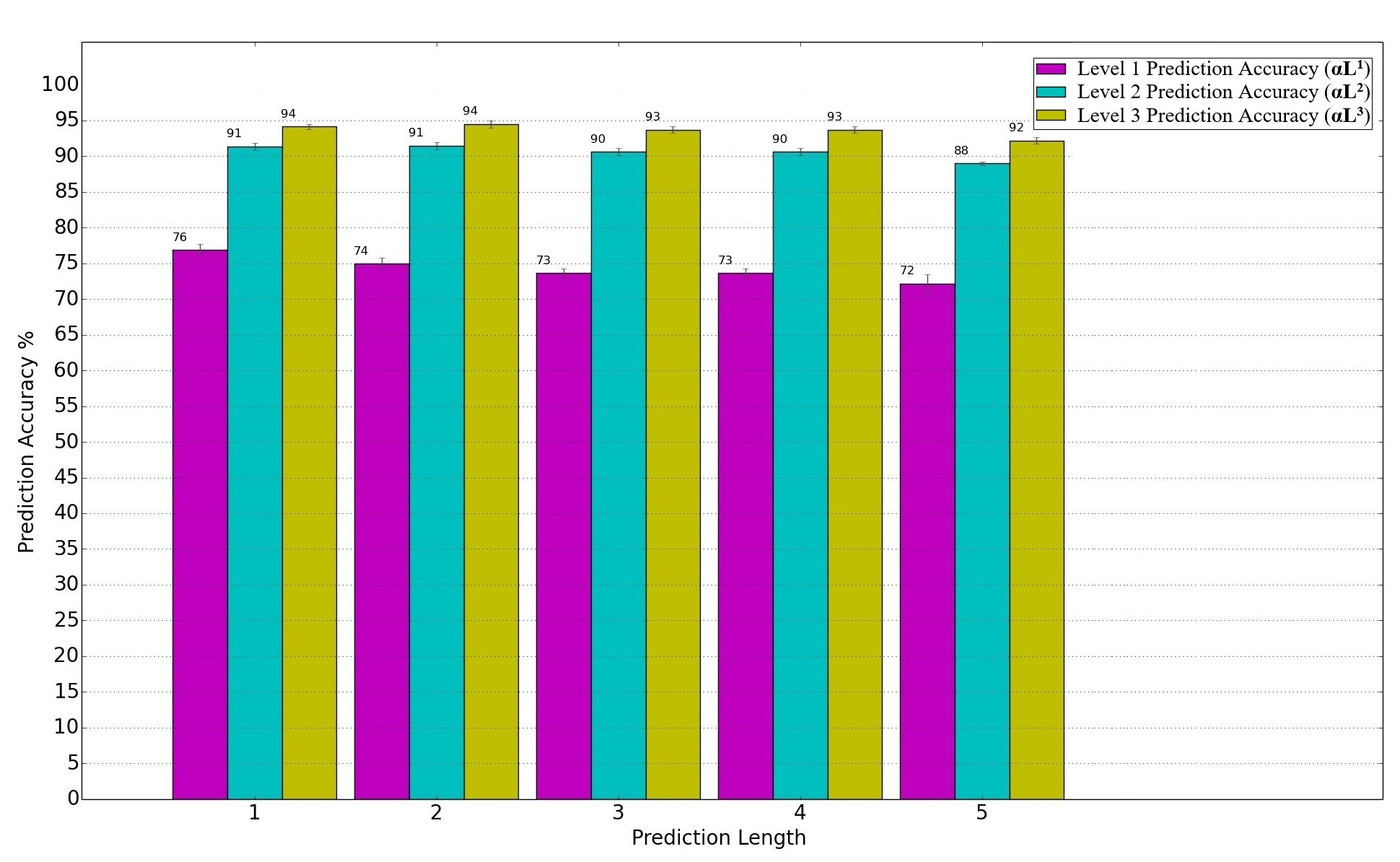}
\caption{Prediction accuracy of alert category variation with multiple data point prediction.}
\label{result12}
\end{figure}

\noindent To examine the multiple data points prediction capability of the HMM prediction module, the length of prediction ($l$) was changed from 1 to 5 and accuracy of prediction was recorded. Accuracy of future alert category predictions is shown in Figure \ref{result12}. With the increment of prediction length, the prediction accuracy was decreased. The maximum prediction accuracies for ${\alpha L}^1$, ${\alpha L}^2$ and ${\alpha L}^3$ were observed as 76\%, 91\% and 94\% correspondingly for single intrusion prediction while the lowest was observed for 5 intrusions prediction as 72\%, 88\% and 92\% respectively.
Fava et al. \cite{fava2008projecting} presented a Cyberattacks projecting method using variable-length markov models. which is the most similar work related to our work. The major differences between their approach and the proposed algorithm is that they have generated three separate Markov models for alert category, alert description and destination IP address attributes of an alert. By using these models, they predicted the next alert category, next alert description and destination IP address separately. With this approach, they achieved an accuracy of 90\% for ${\alpha L}^3$ for predicting the next alert category. When the prediction algorithm proposed in this paper was constrained to only predict the alert category without additional information, it obtained an accuracy of 95\% for ${\alpha L}^3$. Following are the key differences between the proposed algorithm and that of Fava et al. method:
\begin{enumerate}
	\item {Fava et al. used a different dataset for their experiments whereas this paper presents results from the DARPA data set.}
	\item {The closest result that can be compared is the prediction of next alert category. The DARPA data set has nine unique alert categories whereas Fava et al. doesnot disclose the number of unique alert types that present in dataset that they used.}
	\item {When the proposed alert prediction framework was used to predict the alert category as well as the source IP address, the destination IP address, and the alert type, it achieved a level 3 prediction accuracy (${\alpha L}^3$) of 81\%, 81\%, 76\%, 78\%, 75\% and 72\% for cluster sizes of 10, 12, 14, 16, 18 and 20 respectively. Even thought the proposed alert cluster prediction achieved lower prediction accuracy compared to Fava et al. method, it provides critical information the intrusion which will be essential for any response.}
\end{enumerate}

\section{Conclusion}
In this paper, a hidden Markov method based alert prediction framework is proposed. Alert clustering is employed to group selected alert attributes together. A given sequence of alerts is converted to a sequence of alert clusters and then a HMM is used to predict future alert clusters based on the input. The proposed algorithm also provides the alert category as well as the source IP address, the destination IP address, and the alert type, which are critical in responding to an intrusion.\\
Based on the results, it is observed that a smaller number of clusters tend to improve prediction accuracy. The maximum value of ${\alpha L}^3$ was 88\% for 5 clusters and the lowest 31\% was recorded for 50 clusters. When the number of clusters are smaller, it results in a smaller set of unique symbols for the HMM model which improves the learning abilities of the HMM model compared to a larger symbol size. However, when the number of clusters are smaller, it will hinder the separation of unique alert types and cause merging of two or more alert types. As an example, in the DARPA data set, there are nine unique alerts categories generated by Snort IDS. Therefore, if the number of clusters are less than nine, then one alert cluster may include more than one alert category which hinders the prediction.\\
Also, the experimental results indicated that when the number of hidden states are lower than the number of observations (i.e. when the number of observations are 10 and the number of hidden states are between 2 to 4), level 1 prediction accuracy (${\alpha L}^1$) is lower compared to higher number of hidden states (i.e. number of hidden states are between 5 to 8). It is observed that the maximum ${\alpha L}^1$ difference was 24\% for these two scenarios. This shows that when the number of hidden states are low, it may not be possible to model the system states changes efficiently because not enough states are available to represent state transition during a multi-stage intrusion scenario.\\
There are still some challenges that need to be addressed in the proposed alert prediction framework. They include increasing the prediction accuracy with the increase of cluster size and predicting intrusion types that not present in the training data set. Currently it is not possible to predict an intrusion type if that not present in the training data set. Another challenge is identifying false alerts and misleading intrusion actions generated by the attacker in order to mislead intrusion detection systems.


%



\ifCLASSOPTIONcompsoc
  \section*{Acknowledgments}
\else
  \section*{Acknowledgment}
\fi

The support of the NSERC (Natural Sciences and Engineering Research Council of Canada) and University of Western Ontario is gratefully acknowledged by the authors.

\ifCLASSOPTIONcaptionsoff
  \newpage
\fi



\bibliographystyle{IEEEtran}
\bibliography{IEEEabrv,R}

\end{document}